# Electron vortex beams in a magnetic field:
# A new twist on Landau levels and Aharonov-Bohm states

Konstantin Y. Bliokh[1,2], Peter Schattschneider[3,4], Jo Verbeeck[5], and Franco Nori[1,6]

[1]*Advanced Science Institute, RIKEN, Wako-shi, Saitama 351-0198, Japan*
[2]*A. Usikov Institute of Radiophysics and Electronics, NASU, Kharkov 61085, Ukraine*
[3]*Institut für Festkörperphysik, Technische Universität Wien, A-1040 WIEN, Austria*
[4]*University Service Centre for Electron Microscopy, Technische Universität Wien, A-1040 WIEN, Austria*
[5]*EMAT, University of Antwerp, Groenenborgerlaan 171, 2020 Antwerp, Belgium*
[6]*Physics Department, University of Michigan, Ann Arbor, Michigan 48109-1040, USA*

We examine the propagation of the recently-discovered electron vortex beams in a longitudinal magnetic field. We consider both the Aharonov-Bohm configuration with a single flux line and the Landau case of a uniform magnetic field. While stationary Aharonov-Bohm modes represent Bessel beams with flux- and vortex-dependent probability distributions, stationary Landau states manifest themselves as non-diffracting Laguerre-Gaussian beams. Furthermore, the Landau-state beams possess field- and vortex-dependent phases: (i) the Zeeman phase from coupling the quantized angular momentum to the magnetic field and (ii) the Gouy phase, known from optical Laguerre-Gaussian beams. Remarkably, together these phases determine the structure of Landau energy levels. This unified Zeeman-Landau-Gouy phase manifests itself in a nontrivial evolution of images formed by various superpositions of modes. We demonstrate that, depending on the chosen superposition, the image can rotate in a magnetic field with either (i) Larmor, (ii) cyclotron (double-Larmor), or (iii) zero frequency. At the same time, its centroid always follows the classical cyclotron trajectory, in agreement with the Ehrenfest theorem. Remarkably, the non-rotating superpositions reproduce stable multi-vortex configurations that appear in rotating superfluids. Our results open up an avenue for the direct electron-microscopy observation of fundamental properties of free quantum electron states in magnetic fields.

## 1. Introduction

Propagating waves carrying intrinsic orbital angular momentum (OAM), also known as vortex beams, are widely explored and employed in optics [1–3]. These typically represent paraxial wave beams with a "doughnut-like" transverse intensity profile and twisted helical phase. The azimuthal gradient of the helical phase produces a spiraling current and well-defined OAM along the beam axis. Remarkable wave and dynamical features of optical vortices and OAM reveal themselves in the interference between different modes and in interactions with matter.

Few years ago we introduced vortex beams carrying OAM for free quantum electrons [4]. This was followed by several experimental observations using electron microscopy [5–8] and other theoretical investigations [9–14]. The main distinguishing feature of the electron vortex beams, as compared with their optical counterparts, is that they carry a *magnetic moment* proportional to the OAM (up to hundreds [7] of Bohr magnetons per electron!), and, hence, can effectively interact with external *magnetic fields* [4,6,10,12–14]. This feature can be exploited in both free-space fields [4,14] and magnetic structures in solids [6,12,13].

In analogy to magnetic spin properties of quantum electrons, it is natural to expect that the main manifestations of the electron OAM-magnetic field interaction are as follows: (i) Stern-Gerlach-like transport in a transverse field [4], (ii) Larmor precession of the magnetic moment [4,14], and (iii) Zeeman coupling and corresponding energy (phase) shift in a longitudinal field. Note that the case of longitudinal field, which we consider in this paper, is of particular interest because it naturally occurs in electron-microscope lenses. Besides, the longitudinal magnetic field is described by a *vortex* in the vector-potential, and an interesting interplay of coaxial electron and vector-potential vortices appears.

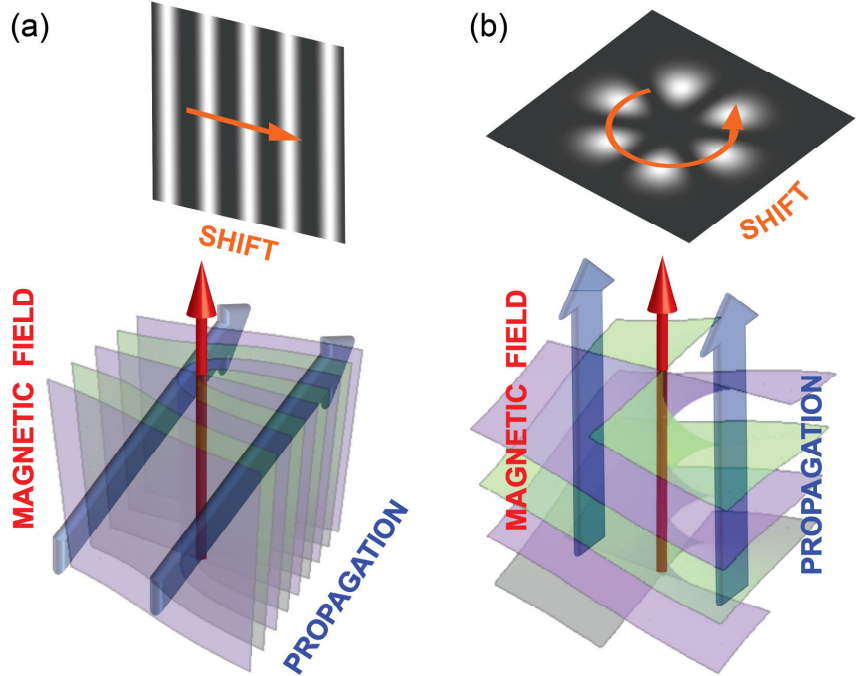

**Fig. 1.** (a) Standard "transverse" geometry of the Aharonov-Bohm and Landau problems. The electron wave propagates *orthogonally* to the magnetic field and shows an *edge* wavefront dislocation in the case of an isolated magnetic flux line. A field-induced transport (e.g., a shift of the interference fringes in the Aharonov-Bohm effect, or quantum-Hall effect) occurs in the direction orthogonal to both the magnetic field and the main propagation of the electrons. (b) In contrast to this, the "longitudinal" geometry, considered this work, involves the propagation of electron waves *along* the magnetic field. In this case, the wavefronts carry a *screw* dislocation, whereas the field-induced transport (e.g., a shift of the Landau interference patterns or radial Aharonov-Bohm effect) occurs in the azimuthal or radial directions (see Figs. 6 and 3 below).

Stationary quantum-electron states in a constant magnetic field are well known as quantized *Landau states* [15,16]. Remarkably, they contain vortex phases and have a well-defined OAM along the magnetic field. However, so far these states were mostly associated with collective properties of condensed-matter electrons inside solids in a magnetic field: such as the quantum-Hall, De Haas-van Alphen, and Shubnikov-De Haas effects [17,18]. In a similar way, electron vortex OAM states naturally occur in the presence of an infinitely thin magnetic solenoid, i.e., as eigenstates of the *Aharonov-Bohm* problem and related problems in superconductors [19–21]. Individual Landau and Aharonov-Bohm states underpin a number of fundamental condensed-matter phenomena, but so far they have never been observed directly in free-space magnetic fields.

In this paper, we examine the propagation of OAM vortex modes in a longitudinal magnetic field. We argue that electron vortex beams allow the *direct observation* and studies of the Aharonov-Bohm and Landau quantum states in magnetic fields. It should be emphasized that in the typical Landau and Aharonov-Bohm problem, the free propagation along the magnetic field is eliminated, while the *transverse* transport and energy levels are considered (Fig. 1). In contrast, in vortex beams, the energy is fixed while the free *longitudinal propagation* unveils new remarkable dynamics, see Fig. 1. We show that the propagating Aharonov-Bohm modes represent *Bessel beams* with field-dependent probability and current distributions, the propagating Landau states represent non-diffracting *Laguerre-Gaussian (LG) beams*, similar to those in multimode optical fibers [22]. Furthermore, the Landau modes acquire nontrivial field- and vortex-dependent phases upon propagation in the magnetic field: (i) the *Zeeman phase* from coupling the quantized OAM to a magnetic field and (ii) the *Gouy phase*, known from optical LG beams [1,23]. Together, these phases determine the structure of *Landau energy levels*. This unified Zeeman-Landau-Gouy phase manifests itself in a nontrivial evolution of interference patterns formed by various superpositions of modes. Namely, the interference patterns rotate in a magnetic field with a rate which is strongly dependent on the chosen superposition of modes – the angular velocity can vary between the Larmor, cyclotron, and zero frequencies. This allows direct experimental observations of different terms in the Landau-level structure, akin to optical pattern rotations in Gouy-phase diffraction experiments [24–27] and Berry-phase [28–32] and rotational-Doppler-effect [33–35] observations. (It is worth noticing that the optical Berry-phase and rotational-Doppler (Coriolis) effects are caused by rotations of the reference frame and are similar to the electron Zeeman effect according to Larmor's theorem.)

## 2. Bessel and Laguerre-Gaussian beams in free space

For completeness and convenience of the exposition below, in this Section we summarize the main properties of the electron vortex beams in free space. To start with, let us recall that the canonical momentum and OAM operators in the coordinate representation are $\hat{\mathbf{p}} = -i\hbar\nabla$ and $\hat{\mathbf{L}} = \mathbf{r} \times \hat{\mathbf{p}}$, respectively. Using cylindrical coordinates $(r, \varphi, z)$, this yields $\hat{L}_z = -i\hbar\partial_\varphi$. The eigenmodes of $\hat{p}_z = -i\hbar\partial_z$ are $\psi \propto \exp(ik_z z)$, i.e., plane waves propagating along the $z$-direction, with continuous-spectrum eigenvalues $p_z = \hbar k_z$. At the same time, the eigenmodes of OAM $\hat{L}_z$ are $\psi_\ell \propto \exp(i\ell\varphi)$, i.e., *vortices*, which are characterized by the quantized eigenvalues $L_z = \hbar\ell$, where $\ell = 0, \pm 1, \pm 2, ...$ is the vortex charge. Modes with well-defined momentum $p_z$ and OAM $L_z$ are called *vortex beams*. Naturally, the expectation value of the canonical OAM (normalized per one particle) for vortex modes is determined as

$$\langle L_z \rangle = \frac{\langle \psi | \hat{L}_z | \psi \rangle}{\langle \psi | \psi \rangle} = \hbar\ell. \qquad (1)$$

Hereafter, the inner product implies the corresponding volume integration. At the same time, the *local* structure of the vortex can be characterized by the probability distribution $\rho$ and current $\mathbf{j}$:

$$\rho = |\psi|^2, \quad \mathbf{j} = \frac{1}{m}\langle \psi | \hat{\mathbf{p}} | \psi \rangle = \frac{\hbar}{m}\mathrm{Im}(\psi^* \nabla \psi). \qquad (2)$$

It is easy to see that for vortex beams with $\psi_\ell \propto \exp(i\ell\varphi + ik_z z)$ Eqs. (2) result in a nonzero azimuthal component of the current:

$$\mathbf{j}_\ell(r) = \frac{\hbar}{m}\left(\frac{\ell}{r}\mathbf{e}_\varphi + k_z \mathbf{e}_z\right)\rho_\ell(r), \qquad (3)$$

where $\mathbf{e}_\varphi$ and $\mathbf{e}_z$ are the unit vectors of the corresponding coordinates, and we neglected the radial component of the current that might appear due to diffraction. The *kinetic* momentum and OAM spatial densities are given by $m\mathbf{j}$ and $m\mathbf{r} \times \mathbf{j}$, respectively, so that the averaged kinetic OAM per one electron, $\langle \mathcal{L}_z \rangle$, can be calculated via the following volume integration:

$$\langle \mathcal{L}_z \rangle = \frac{m\int r j_\varphi dV}{\int \rho\, dV} = \hbar\ell. \qquad (4)$$

In free space, this naturally coincides with Eq. (1), and one can say that it is the circulating azimuthal current that produces the well-defined OAM in the vortex beams.

Let us explicitly summarize the main examples of free-space electron vortex beams. Assuming mono-energetic electrons, the problem is described by the stationary Schrödinger equation:

$$\hat{H}\psi = E\psi, \quad \hat{H} = \frac{\hat{\mathbf{p}}^2}{2m}, \quad (5)$$

Using cylindrical coordinates, Eq. (5) takes the form

$$-\frac{\hbar^2}{2m}\left[\frac{1}{r}\frac{\partial}{\partial r}\left(r\frac{\partial}{\partial r}\right) + \frac{1}{r^2}\frac{\partial^2}{\partial \varphi^2} + \frac{\partial^2}{\partial z^2}\right]\psi = E\psi. \quad (6)$$

The axially-symmetric solutions of Eq. (6) represent *non-diffracting Bessel beams* [36]:

$$\psi_\ell^B \propto J_{|\ell|}(\kappa r)\exp\left[i(\ell\varphi + k_z z)\right]. \quad (7)$$

Here $J_\ell$ is the Bessel function of the first kind, whereas $\kappa$ is the transverse wave number. Solutions (7) satisfy Eq. (6) provided that the following dispersion relation is fulfilled:

$$E = \frac{\hbar^2}{2m}k^2 = \frac{\hbar^2}{2m}\left(k_z^2 + \kappa^2\right). \quad (8)$$

Evidently, the Bessel beams (7) are vortex beams, i.e., eigenmodes of momentum $\hat{p}_z$ and OAM $\hat{L}_z$ with the corresponding eigenvalues $p_z = \hbar k_z$ and $L_z = \hbar\ell$. Calculating the probability density and current (2) for modes (7), we obtain:

$$\rho_{|\ell|}^B(r) \propto \left|J_{|\ell|}(\kappa r)\right|^2, \quad \mathbf{j}_\ell^B(r) = \frac{\hbar}{m}\left(\frac{\ell}{r}\mathbf{e}_\varphi + k_z \mathbf{e}_z\right)\rho_{|\ell|}^B(r). \quad (9)$$

These transverse distributions for Bessel beams with different OAM quantum numbers $\ell$ are shown in Figure 2. Substituting the wave function (7) and probability distributions (9) into Eqs. (1) and (4), it is easy to verify that the OAM expectation values are $\langle L_z\rangle = \langle \mathcal{L}_z\rangle = \hbar\ell$.

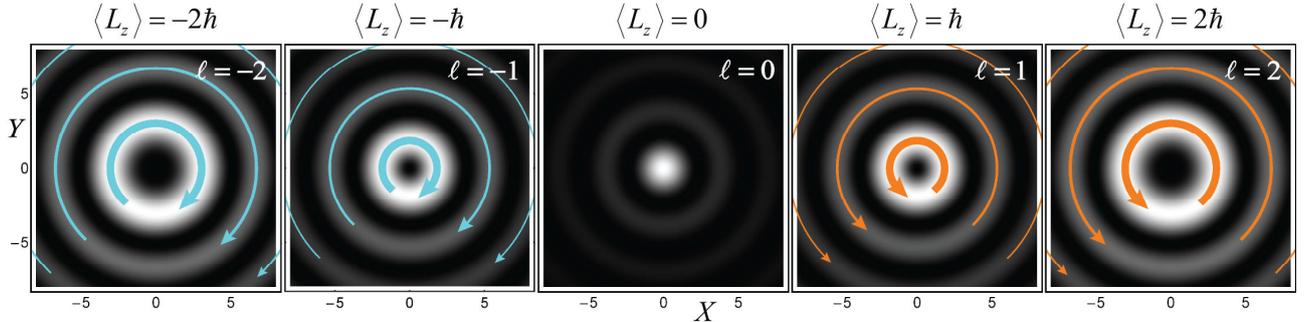

**Fig. 2.** Transverse probability density distributions and azimuthal currents, Eqs. (9), for free-space Bessel beams (7) with different OAM quantum numbers $\ell$. The dimensionless coordinates $X = \kappa x$ and $Y = \kappa y$ are used. Here and in the following figures, the radii and thicknesses of the current circles correspond to the positions and values of the maxima (normalized in each panel independently) of the quantity $r j_\varphi$ that determines the contribution to the kinetic OAM, Eq. (4).

Bessel beams have a simple mathematical form and are well suited for systems with radially-limited apertures. The Bessel modes in such systems have been described elsewhere [9]. However, in unbounded free space they are not well localized in the radial direction – because the integral $\int \rho_{|\ell|}^B r\, dr$ is divergent. The simplest transversely-confined vortex beams in free space are *diffracting LG beams* [1]. They are solutions of the approximate paraxial Schrödinger equation (6) with $\partial^2/\partial z^2 \simeq k^2 + 2ik\partial/\partial z$, where it is assumed that $k^2 - k_z^2 \ll k^2$, i.e., the transverse wave number is small. In this approximation, the LG beams have the form [1]

$$\psi_{\ell,n}^{LG} \propto \left(\frac{r}{w(z)}\right)^{|\ell|} L_n^{|\ell|}\left(\frac{2r^2}{w^2(z)}\right) \exp\left(-\frac{r^2}{w^2(z)} + ik\frac{r^2}{2R(z)}\right) e^{i(\ell\varphi + kz)} e^{-i(2n+|\ell|+1)\zeta(z)}. \qquad (10)$$

where $L_n^{|\ell|}$ are the generalized Laguerre polynomials, $n = 0, 1, 2, \ldots$ is the radial quantum number, $w(z) = w_0\sqrt{1 + z^2/z_R^2}$ is the beam width, which depends on $z$ due to diffraction, $R(z) = z\left(1 + z_R^2/z^2\right)$ is the radius of curvature of the wavefronts, and $\zeta(z) = \arctan(z/z_R)$. Here the characteristic transverse and longitudinal scales of the beam are the waist $w_0$ (width in the focal plane $z = 0$) and the Rayleigh diffraction length $z_R$:

$$w_0 \gg 2\pi/k^{-1}, \quad z_R = kw_0^2/2 \gg w_0. \qquad (11)$$

Note that the last exponent factor in Eq. (10) describes the *Guoy phase* [23] – it yields an additional phase difference

$$\Phi_G = (2n + |\ell| + 1)\pi \qquad (12)$$

upon the beam propagation from $z/z_R \ll -1$ to $z/z_R \gg 1$. The Gouy phase is closely related to the *transverse confinement* of the modes [37]. The dispersion relation for LG beams is simply $E = \hbar^2 k^2/2m$ [cf. Eq. (8)], whereas the small transverse wave-vector components are taken into account in the $z$-dependent diffraction terms.

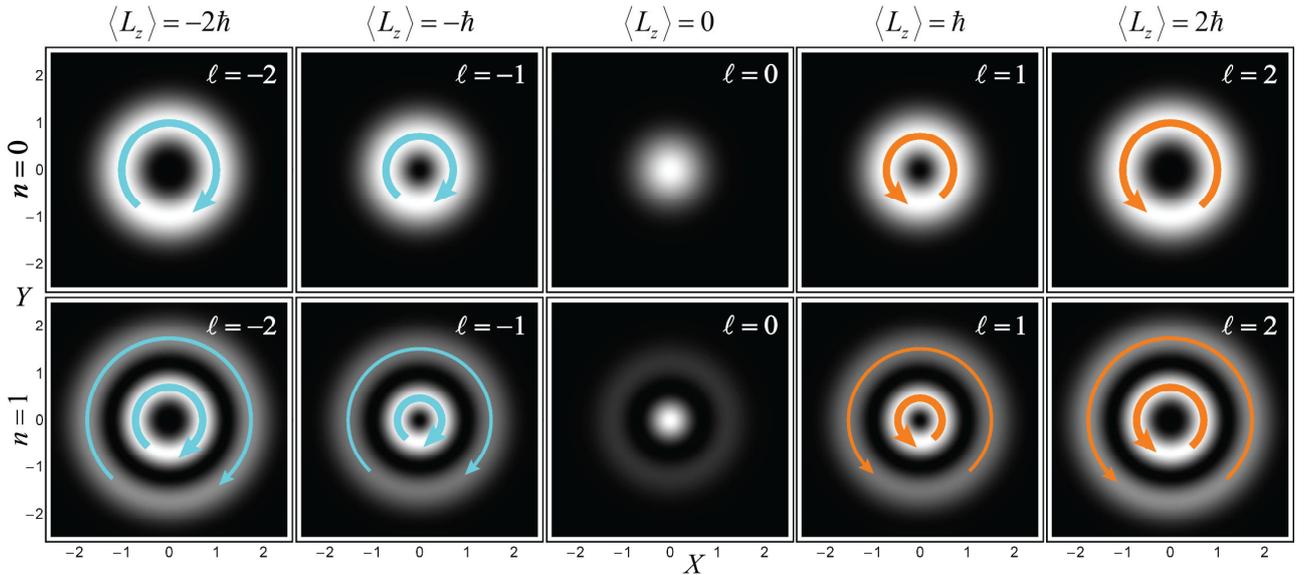

**Fig. 3.** Transverse probability density distributions and azimuthal currents, Eqs. (13), for free-space LG beams (10) at $z = 0$ with different OAM quantum numbers $\ell$ and radial indices $n = 0$ (upper row) and $n = 1$ (lower row). The dimensionless coordinates $X = x/w_0$ and $Y = y/w_0$ are used.

Clearly, the LG beams (10) are eigenmodes of the OAM $\hat{L}_z$, with eigenvalues $L_z = \hbar\ell$. At the same time, they are *approximate* eigenmodes of $\hat{p}_z$ with eigenvalues $p_z \simeq \hbar k$, as long as the diffraction effects are neglected. Calculating the probability density and the azimuthal component of the current (2) for the modes (10), we obtain:

$$\rho_{|\ell|,n}^{LG}(r,z) \propto \left(\frac{r^2}{w^2(z)}\right)^{|\ell|} \left|L_n^{|\ell|}\left(\frac{2r^2}{w^2(z)}\right)\right|^2 \exp\left(-\frac{2r^2}{w^2(z)}\right), \quad j_{\ell,n\,\varphi}^{LG}(r,z) = \ell\frac{\hbar}{mr}\rho_{|\ell|,n}^{LG}(r,z), \qquad (13)$$

Figure 3 shows these transverse distributions at $z = 0$ for the LG modes with different quantum numbers $\ell$ and $n$. It is easy to see that the OAM expectation values, Eqs. (1) and (4), are $\langle L_z \rangle = \langle \mathcal{L}_z \rangle = \hbar \ell$.

Up to this point, the properties of the electron vortex beams are entirely analogous to those of the corresponding optical beams in free space [1–3], but the presence of a magnetic field and the electric charge of the electron introduce an interaction which has no optical counterpart.

## 3. Bessel beams in the presence of an Aharonov-Bohm flux

Prior to proceed with explicit examples, we describe the basic fundamental features of the electron OAM in the presence of a magnetic field $\mathbf{B}(\mathbf{r})$. This problem is described by the Schrödinger equation (5) with modified *kinetic* (or *covariant*) momentum $\hat{\boldsymbol{p}} = \hat{\mathbf{p}} - e\mathbf{A}$:

$$\hat{H} = \frac{\hat{\boldsymbol{p}}^2}{2m} = \frac{1}{2m}(\hat{\mathbf{p}} - e\mathbf{A})^2, \tag{14}$$

where $\mathbf{A}(\mathbf{r})$ is the magnetic vector-potential, $\mathbf{B} = \nabla \times \mathbf{A}$, and $e = -|e|$ is the electron charge. The kinetic momentum characterizes the velocity and mechanical momentum of the electron; it also determines the probability current satisfying the continuity equation [15]:

$$\rho = |\psi|^2, \quad \mathbf{j} = \frac{1}{m}\langle\psi|\hat{\boldsymbol{p}}|\psi\rangle = \frac{\hbar}{m}\mathrm{Im}(\psi^*\nabla\psi) - \frac{e}{m}\mathbf{A}\rho. \tag{15}$$

Using kinetic momentum $\hat{\boldsymbol{p}}$, one can define the operator of *kinetic OAM*, $\hat{\mathcal{L}} = \mathbf{r} \times \hat{\boldsymbol{p}}$, which yields $\hat{\mathcal{L}}_z = -i\hbar\partial_\varphi - erA_\varphi$ [38,39]. The circulation of the probability current (15) yields the expectation value of the kinetic OAM, cf. Eqs. (1) and (4):

$$\langle \mathcal{L}_z \rangle = \frac{\langle\psi|\hat{\mathcal{L}}_z|\psi\rangle}{\langle\psi|\psi\rangle} = \frac{m\int r j_\varphi dV}{\int \rho\, dV}. \tag{16}$$

Note that the kinetic quantities (15) and (16) are invariant under the gauge transformation

$$\mathbf{A} \to \mathbf{A} + \nabla\chi, \quad \psi \to \psi \exp\left(i\frac{e}{\hbar}\chi\right), \tag{17}$$

[$\chi(\mathbf{r})$ is an arbitrary function], and they describe the observable mechanical OAM of the electron.

Thus, in the presence of a magnetic field, the canonical and kinetic OAM, Eqs. (1) and (16), differ from each other. Noteworthy, in the case of an axially-symmetric longitudinal magnetic field, the vector-potential can be chosen in the form of a *vortex*, $\mathbf{A}(\mathbf{r}) = A(r)\mathbf{e}_\varphi$, and the two summands in the second Eq. (15) represent either *co*-circulating or *counter*-circulating vortices for vortex beams. Hence, depending on the mutual direction of the OAM and magnetic field, the kinetic OAM (16) is either *enhanced* or *diminished* by the magnetic vector-potential. As we show below, this effect occurs even in the Aharonov-Bohm configuration, i.e., with the electron wave-function localized outside the magnetic field.

Let us now consider the electron vortex beams in the presence of a magnetic flux line (i.e., an infinitely thin shielded solenoid) directed along the $z$-axis and containing the flux $\phi$. Such magnetic field is described by the vortex vector-potential

$$\mathbf{A}(\mathbf{r}) = \frac{\phi}{2\pi r}\mathbf{e}_\varphi, \tag{18}$$

and the Schrödinger equation (14) in cylindrical coordinates takes the form

$$-\frac{\hbar^2}{2m}\left[\frac{1}{r}\frac{\partial}{\partial r}\left(r\frac{\partial}{\partial r}\right) + \frac{1}{r^2}\left(\frac{\partial}{\partial \varphi} - i\alpha\right)^2 + \frac{\partial^2}{\partial z^2}\right]\psi = E\psi. \tag{19}$$

Here $\alpha = \dfrac{e\phi}{2\pi\hbar}$ is the dimensionless magnetic-flux parameter.

Aharonov and Bohm found cylindrical eigenmodes of Eq. (19) [19], which, in fact, represent *Bessel beams* (7) with the Bessel-function order *shifted* by the magnetic flux parameter $\alpha$ (see Fig. 4):

$$\psi_\ell^{AB} \propto J_{|\ell-\alpha|}(\kappa r)\exp\left[i(\ell\varphi + k_z z)\right]. \tag{20}$$

Here the wave numbers still satisfy the dispersion relation (8). Equation (20) is also consistent with Berry's prescription [20] concerning the shift of the order of the radial function of the OAM modes. Note that the magnetic flux parameter $\alpha$ can be regarded as a *vortex charge of the vector-potential* $\mathbf{A}$, i.e., the Dirac phase [40] calculated for the closed loop $C = \{r = \mathrm{const}, \varphi \in (0, 2\pi)\}$ and divided by $2\pi$:

$$\Phi_D = \frac{e}{\hbar}\oint_C \mathbf{A}\cdot d\mathbf{r} = \frac{e\phi}{\hbar} = 2\pi\alpha. \tag{21}$$

Evidently, the modified Bessel beams (20) are still eigenmodes of the *canonical* OAM operator $\hat{L}_z$, with the eigenvalues $L_z = \hbar\ell$ and expectation value (1) $\langle L_z \rangle = \hbar\ell$. At the same time, the kinetic OAM (16) is different. Calculating the probability density and current (15), for the Aharonov-Bohm states (20), we arrive at

$$\rho^{AB}_{|\ell-\alpha|}(r) \propto \left|J_{|\ell-\alpha|}(\kappa r)\right|^2, \quad \mathbf{j}^{AB}_{\ell-\alpha}(r) = \frac{\hbar}{m}\left(\frac{\ell-\alpha}{r}\mathbf{e}_\varphi + k_z \mathbf{e}_z\right)\rho^{AB}_{|\ell-\alpha|}(r). \tag{22}$$

Substituting this into Eq. (16), the expectation value of the kinetic OAM yields [38,39]

$$\langle \mathcal{L}_z \rangle = \hbar(\ell - \alpha). \tag{23}$$

Thus, as predicted, depending on the sign of $\alpha$, the magnetic flux can either enhance or diminish the kinetic OAM, compared to the canonical OAM value $\hbar\ell$.

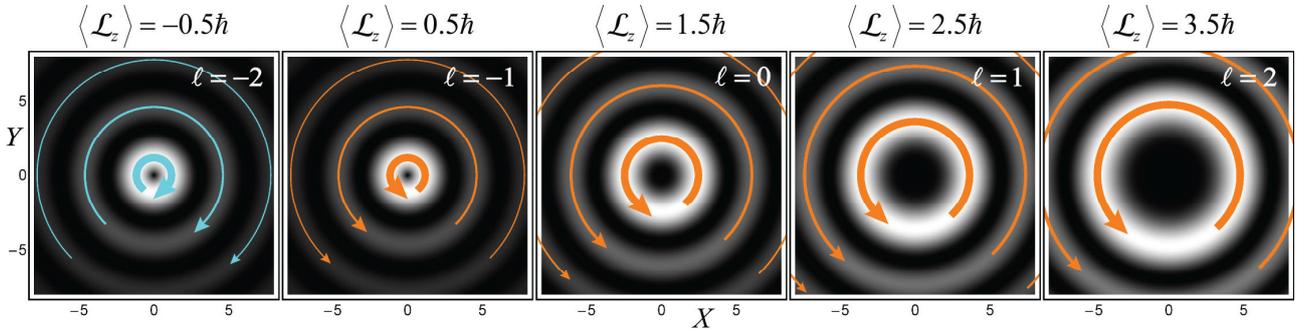

**Fig. 4.** Transverse probability density distributions and azimuthal currents, Eqs. (22), for the Aharonov-Bohm Bessel modes (20) with different OAM quantum numbers $\ell$ in the presence of a magnetic flux line of strength $\alpha = -1.5$. The free-space symmetry between modes with the opposite $\ell$ (see Fig. 2) is significantly broken here by the magnetic flux $\alpha$. The probability currents generate kinetic OAM (23), which is shown above the panels. The dimensionless coordinates $X = \kappa x$ and $Y = \kappa y$ are used.

Equations (22) and (23) reveal an important general feature of the OAM solutions in the presence of a magnetic field. Namely, the vector-potential contribution *breaks the symmetry* between the current distributions and OAM in the modes with opposite vorticities $\pm\ell$ or opposite magnetic fields $\pm\alpha$. The observables $\rho$, $|j_\varphi|$ and $|\langle \mathcal{L}_z \rangle|$ are invariant with respect to the transformation $(\ell,\alpha) \to (-\ell,-\alpha)$, but neither with respect to $(\ell,\alpha) \to (-\ell,\alpha)$ nor $(\ell,\alpha) \to (\ell,-\alpha)$. This resembles the effect of spin-orbit interaction considered in [10,41], but here it is caused by the *Zeeman-type interaction* between OAM and magnetic field. We do not see any *Zeeman energy* or *phase* in the Aharonov-Bohm modes, because, in the case of the magnetic flux line, the wave function (20) with $\langle \mathcal{L}_z \rangle \neq 0$ is localized *outside* the area of magnetic field. Nonetheless, the Zeeman-type symmetry is present in the probability distributions (22) and kinetic OAM (23) owing to the vector-potential (Aharonov-Bohm) effect. The possibility of the radial Aharonov-Bohm effect in the longitudinal geometry with a vortex beam is also mentioned in [14]. Radial dependencies of the probability densities and azimuthal currents (22) for different Bessel beams (20) are shown in Figure 4. It is seen that the absolute values of the azimuthal current and kinetic OAM are larger for parallel OAM and magnetic field ($-\alpha\ell > 0$) as compared to the case of anti-parallel OAM and field ($-\alpha\ell < 0$). A similar symmetry breaking is also observed in magnetic electronic transitions that underpin the asymmetric scattering of electron vortex beams on magnetized samples [6].

It is worth remarking that the absolute value of the kinetic OAM (23) determines the radius of the Bessel beam in the Aharonov-Bohm problem. Namely, the radius of the cylindrical caustic underlying the first radial maximum of the mode, $R_{|\ell-\alpha|}$, can be written as

$$\kappa R_{|\ell-\alpha|} = \hbar^{-1}\left|\langle \mathcal{L}_z \rangle\right| = |\ell - \alpha| = \left|\ell - \frac{\Phi_D}{2\pi}\right|. \quad (24)$$

This is entirely analogous to the quantization of caustics described for the Bessel beams with spin-orbit interaction [10,41]. In this manner, the Dirac phase from the vector potential substitutes here the Berry phase from nonparaxial spins. Note that, in contrast to the Berry phase, the Dirac phase (21) can be arbitrarily large, and here it is clearly observable in its full value, not only modulo $2\pi$.

## 4. Landau levels and Laguerre-Gaussian beams in a uniform magnetic field

We are now in a position to consider electron vortex beams in a uniform magnetic field $\mathbf{B} = B\mathbf{e}_z$. The problem is described by the Hamiltonian (14) with the vortex vector potential

$$\mathbf{A} = \frac{Br}{2}\mathbf{e}_\varphi. \quad (25)$$

This yields the Schrödinger equation

$$-\frac{\hbar^2}{2m}\left[\frac{1}{r}\frac{\partial}{\partial r}\left(r\frac{\partial}{\partial r}\right) + \frac{1}{r^2}\left(\frac{\partial}{\partial \varphi} + i\sigma\frac{2r^2}{w_m^2}\right)^2 + \frac{\partial^2}{\partial z^2}\right]\psi = E\psi, \quad (26)$$

where $w_m = 2\sqrt{\hbar/|eB|} = \sqrt{2\hbar/m|\Omega|}$ is the magnetic length parameter, $\Omega = eB/2m$ is the Larmor frequency corresponding to the $g$-factor $g=1$ for OAM [4,10,14], and we introduced the parameter $\sigma = \text{sgn}\, B = \pm 1$, which indicates the direction of the magnetic field. Note that it is the Larmor frequency $\Omega$, rather than the cyclotron frequency $\omega_c = eB/m = 2\Omega$, which is the fundamental frequency in this problem [14,42]. This is related to the Larmor's theorem, conservation of the angular momentum, and it will be clearly seen below from the quantum picture of the electron evolution. Solutions of Eq. (26) with well defined OAM are known as quantized *Landau states* [15,16,43]. Remarkably, they have the form of *non-diffracting LG beams* (see Fig. 5):

$$\psi_{\ell,n}^L \propto \left(\frac{r}{w_m}\right)^{|\ell|} L_n^{|\ell|}\left(\frac{2r^2}{w_m^2}\right)\exp\left(-\frac{r^2}{w_m^2}\right)\exp\left[i(\ell\varphi + k_z z)\right]. \quad (27)$$

The Landau states (27) are identical to the LG beams (10) with beam waist $w_0 = w_m$ at $z = 0$. We also introduce a longitudinal scale $z_m$ determined by the Larmor frequency and the electron velocity $v = \sqrt{2E/m}$: $z_m = v/|\Omega|$. The transverse magnetic length $w_m$ and longitudinal Larmor length $z_m$ represent counterparts of the beam waist and Rayleigh length of the free-space LG beams, Eqs. (11), but here they are uniquely determined by the electron energy and magnetic field strength:

$$w_m = \frac{2\sqrt{\hbar}}{\sqrt{|eB|}}, \quad z_m = \frac{2\sqrt{2Em}}{|eB|}, \quad \text{i.e.,} \quad z_m = \sqrt{\frac{E}{\hbar|\Omega|}} w_m. \quad (28)$$

In optics, non-diffracting LG modes entirely analogous to Eq. (27) appear in parabolic-index optical fibers [22]. This is related to the fact that the Schrödinger equation in a uniform magnetic field can be mapped onto a two-dimensional quantum-oscillator problem [16,14], as well as paraxial focused beams [42]. The fact that non-diffracting eigenmodes in the magnetic field are transversely confined (i.e., possess a discrete radial quantum number $n$) reflects the boundedness of classical electron orbits in magnetic field.

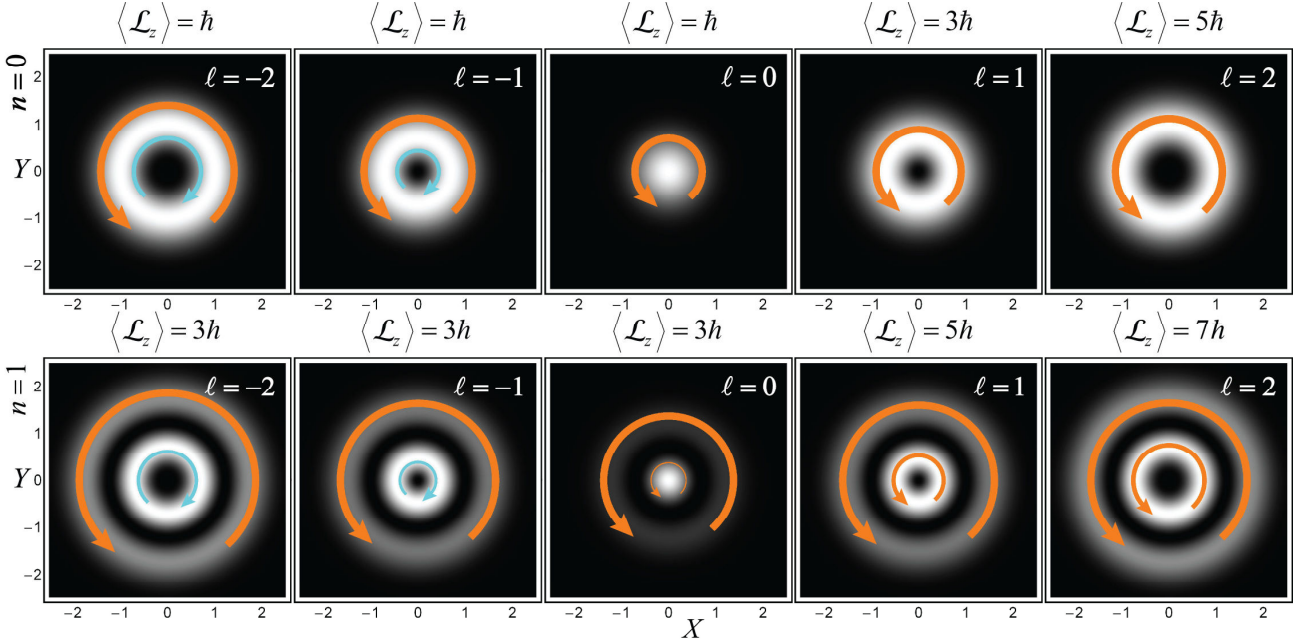

**Fig. 5.** Transverse probability density distributions and azimuthal currents, Eqs. (30), for the Landau LG modes (27) with different OAM quantum numbers $\ell$ in magnetic field with $\sigma = 1$. While the probability density distributions have the same form as for the LG beams in free space (see Fig. 3), the current distributions differ drastically due to the strong vector-potential contribution. This makes the integral azimuthal current and kinetic OAM (33) always positive, independently of the sign of $\ell$. The dimensionless coordinates $X = x/w_m$ and $Y = y/w_m$ are used.

While the diffracting LG beams (10) represent an *approximate* paraxial solutions of the Schrodinger equation, Landau LG modes (27) yield *exact* solutions of the problem with magnetic field. In doing so, the wave numbers satisfy the following dispersion relation:

$$E = \frac{\hbar^2 k_z^2}{2m} - \hbar\Omega\ell + \hbar|\Omega|(2n+|\ell|+1) \equiv E_\parallel + \underbrace{E_Z + E_G}_{E_\perp}. \quad (29)$$

While $E_\parallel = \hbar^2 k_z^2 / 2m$ is the energy of the free longitudinal motion, the quantized transverse-motion energy in Eq. (29) can be written as

$$E_\perp = \hbar|\Omega|(2N+1), \quad \text{with} \quad N = n + |\ell|[1+\text{sgn}(\sigma\ell)]/2 = 0,1,2,... \quad (29')$$

Thus, Eq. (29') describes the structure of *Landau energy levels* [15,16]. In the form of Eq. (29), we see that the term $E_Z = -\hbar\Omega\ell = \mu_B \ell B$ ($\mu_B = |e|\hbar/2m$) represents the *Zeeman energy* of the canonical OAM $\hbar\ell$ in a magnetic field $B$. At the same time, the term $E_G = \hbar|\Omega|(2n+|\ell|+1)$ can be associated with the *Gouy phase* (12) of the diffractive LG modes. (The Gouy-phase term can be associated with the transverse kinetic energy of spatially-confined modes, which shifts the propagation constants and eigenfrequencies of the waveguide and resonator modes [23,37].) Thus, one can say that the Landau energy of an electron in a magnetic field is the sum of the Zeeman and Gouy contributions: $E_\perp = E_Z + E_G$. This is one of the central points in this paper. As we show below, these two contributions are separately observable and lead to remarkable behaviour of interference patterns in a magnetic field.

Obviously, the Landau LG modes (27) are eigenmodes of the momentum $\hat{p}_z$ and the canonical OAM $\hat{L}_z$ with corresponding eigenvalues $p_z = \hbar k_z$ and $L_z = \hbar \ell$, and the expectation value (1) $\langle L_z \rangle = \hbar \ell$. Calculating the probability density and current (15), we obtain

$$\rho_{|\ell|,n}^L(r) \propto \left(\frac{r^2}{w_m^2}\right)^{|\ell|} \left|L_n^{|\ell|}\left(\frac{2r^2}{w_m^2}\right)\right|^2 \exp\left(-\frac{2r^2}{w_m^2}\right),$$

$$\mathbf{j}_{\ell,n}^L(r) = \frac{\hbar}{m}\left[\frac{1}{r}\left(\ell + \sigma \frac{2r^2}{w_m^2}\right)\mathbf{e}_\varphi + k_z \mathbf{e}_z\right]\rho_{|\ell|,n}^L(r), \tag{30}$$

It is worth noticing that for counter-circulating vortex $\exp(i\ell\varphi)$ and vector-potential $A_\varphi$, $\ell\sigma < 0$, the azimuthal current changes sign at $r = r_{|\ell|} \equiv |\ell| w_m / \sqrt{2}$, i.e., around the first radial maximum of the LG mode. For $r < r_{|\ell|}$ the current from the vortex $\exp(i\ell\varphi)$ prevails, whereas for $r > r_{|\ell|}$ the contribution from the vector-potential $A_\varphi$ becomes dominant (see Fig. 5). When integrated, the vector-potential contribution always exceeds the vortex one. Indeed, calculating the kinetic OAM (16) for the Landau state (27) we have:

$$\langle \mathcal{L}_z \rangle = \hbar\left(\ell + \sigma\left\langle\frac{2r^2}{w_m^2}\right\rangle\right), \tag{31}$$

Here the quantity

$$\left\langle\frac{2r^2}{w_m^2}\right\rangle = \frac{\langle\psi|2r^2/w_m^2|\psi\rangle}{\langle\psi|\psi\rangle} = 2\frac{\int_0^\infty \left[\xi^{|\ell|} L_n^{|\ell|}(2\xi^2) e^{-\xi^2}\right]^2 \xi^3 d\xi}{\int_0^\infty \left[\xi^{|\ell|} L_n^{|\ell|}(2\xi^2) e^{-\xi^2}\right]^2 \xi d\xi} = (2n + |\ell| + 1) \tag{32}$$

determines the squared "spot size" of the LG beam [45]. Hence, the magnetic-field correction to the kinetic OAM is directly related to the Gouy energy of the LG modes, Eq. (29), and the resulting OAM (31) becomes

$$\langle \mathcal{L}_z \rangle = \hbar\left[\ell + \sigma(2n + |\ell| + 1)\right]. \tag{33}$$

Using Eq. (29'), this equation can also be written as

$$\langle \mathcal{L}_z \rangle = \hbar\sigma(2N + 1), \quad N = n + |\ell|[1 + \text{sgn}(\sigma\ell)]/2 = 0, 1, 2, \ldots \tag{33'}$$

Hence, the sign of the kinetic OAM is solely determined by the direction of the magnetic field, $\sigma$, and is independent of the vortex charge $\ell$. Note that the value of $\langle \mathcal{L}_z \rangle$ is independent of the magnitude of the magnetic field, $|B|$, because the radius of the beam changes as $w_m \propto 1/\sqrt{|B|}$, Eq. (28), whereas the angular velocity $\Omega \propto |B|$, and $L_z \propto \Omega w_m^2$. In contrast to classical electron motion in a magnetic field, which can have zero OAM, Eq. (33') shows that there is a *minimal kinetic OAM* of quantum Landau states: $|\langle \mathcal{L}_z \rangle|_{\min} = \hbar$. Note also that for parallel OAM and magnetic field, $\sigma\ell > 0$, the canonical OAM $\hbar\ell$ is enhanced (in absolute value) by the magnetic-field contribution: $\langle \mathcal{L}_z^> \rangle = \hbar[2\ell + \sigma(2n + 1)]$. At the same time, in the opposite case of anti-parallel

OAM and magnetic field, $\sigma\ell < 0$, the kinetic OAM takes the form $\langle \mathcal{L}_z^< \rangle = \hbar\sigma(2n+1)$, i.e., becomes *independent* of the vortex charge $\ell$. This is caused by the partial cancellation of the counter-circulating azimuthal currents produced by the vortex $\exp(i\ell\varphi)$ and by the magnetic vector-potential $A_\varphi$.

Using the kinetic OAM (33), the Landau level structure (29) can be written as

$$E = \frac{\hbar^2 k_z^2}{2m} - \hbar\Omega\langle \mathcal{L}_z \rangle. \tag{34}$$

In other words, the Landau levels can be described by a single Zeeman energy, whereas the Gouy term is incorporated into the kinetic OAM. This is consistent with the actual magnetic moment of the electron in the presence of a magnetic field. Indeed, the magnetic moment can be defined using the current as [10]

$$\langle \mathcal{M} \rangle = \frac{e}{2} \frac{\int (\mathbf{r}\times\mathbf{j})\,dV}{\int \rho\,dV} = \frac{e}{2m}\langle \mathcal{L} \rangle, \tag{35}$$

so that the Zeeman energy in Eq. (34) is equal to $E_\perp = -\langle \mathcal{M}_z \rangle B$.

Figure 5 shows transverse distributions of the probability densities and currents (30), as well as kinetic OAM (33), for different Landau LG beams (27). Akin to the Aharonov-Bohm Bessel beams, these demonstrate the asymmetry of the azimuthal currents $|j_\varphi|$ and kinetic OAMs $|\langle \mathcal{L}_z \rangle|$ for the modes with opposite vorticities $\pm\ell$ or opposite magnetic-field directions $\pm\sigma$. However, unlike the Aharonov-Bohm case, the probability distribution $\rho$ in Eq. (30) is independent of $\mathrm{sgn}\,\ell$ and $\sigma$. Correspondingly, the radii of the Landau LG modes are determined by the absolute value of the OAM quantum number, $|\ell|$, and radial quantum number $n$ [cf. Eq. (24)]. In contrast to the Aharonov-Bohm modes with common field-independent dispersion (8), for Landau modes the main features of the interactions with the magnetic field are contained in the dispersion relations (29) and (34). These Landau-Zeeman-Gouy relations bring about phases strongly dependent on the mode quantum numbers and the magnetic field, and below we argue that this can be employed in interferometry of electron vortex beams.

## 5. Unveiling Landau-Zeeman-Gouy phases via image rotations

In the typical Landau-level problems, the $z$-propagation is eliminated and the quantized energy levels (29) underlie the transverse electron transport [17,18]. In contrast, here we consider the LG vortex modes (27) and their superpositions generated in a system with a *fixed* electron energy $E$ and *free propagation* along the $z$-axis. Then, different modes will have different longitudinal wave numbers $k_z$ satisfying the dispersion relation (29) and (34). Assuming paraxial electrons, $E_\perp \ll E$, $w_m \ll z_m$, one can represent the wave number as $k_z \simeq k + \Delta k_z$, where $\hbar k = \sqrt{2Em}$ and

$$\Delta k_z = -\left[\sigma\ell + (2n+|\ell|+1)\right]/z_m. \tag{36}$$

It is seen from here that the Larmor length $z_m$, Eq. (28), determines the characteristic longitudinal scale of the beam evolution. Upon propagation along the $z$-axis, the magnetic-field correction to the longitudinal wave vector, Eq. (36), yields an additional phase

$$\Phi_{LZG} = \Delta k_z z. \tag{37}$$

Equations (36) and (37) describe the vortex- and field-dependent phase, which we call the *Landau-Zeeman-Gouy phase* because of its intimate relation to the Landau levels, Zeeman coupling, and Gouy phase. This phase is one of the central results of this paper, which reveals itself in a rich evolution of interference patterns in a magnetic field.

Let us first consider an "OAM-balanced" superposition of two LG modes (27) with the same radial number $n$ but opposite vortex charges $\pm\ell$:

$$\psi^{(1)} = \psi^L_{-\ell,n} + \psi^L_{\ell,n}. \tag{38}$$

This superposition has zero net canonical OAM, $\langle L_z \rangle = 0$, and is characterized by a "flower-like" symmetric pattern with $2|\ell|$ radial petals [24,46]: $|\psi^{(1)}|^2 \propto \cos^2(\ell\varphi)$ (see Fig. 6). The difference of phases (37) for the modes $\psi^L_{\pm\ell,n}$ is described by the Zeeman terms $\Delta\Phi^{(1)}_{LZG} = \mp\ell\sigma\Phi_0$, $\Phi_0 = z/z_m$. These phases modify the azimuthal vortex dependencies as $\exp[\pm i\ell\varphi] \to \exp[\pm i\ell(\varphi - \sigma\Phi_0)]$ and results in the *rotation of the interference pattern* by the angle

$$\Delta\varphi = \sigma z / z_m. \tag{39}$$

Since $z/z_m = |\Omega| z/v$, the rotation (39) is characterized by the *Larmor frequency* $|\Omega|$. Examples of the Larmor-frequency rotations (39) are shown in Fig. 6. Mathematically, this rotation is analogous to the optical image rotations caused by the Berry phase [28–32] or rotational-Doppler (Coriolis) effect [33–35]. Indeed, all these phenomena are described by $\ell$-dependent phases, which are similar in the rotational and magnetic-field systems according to Larmor's theorem.

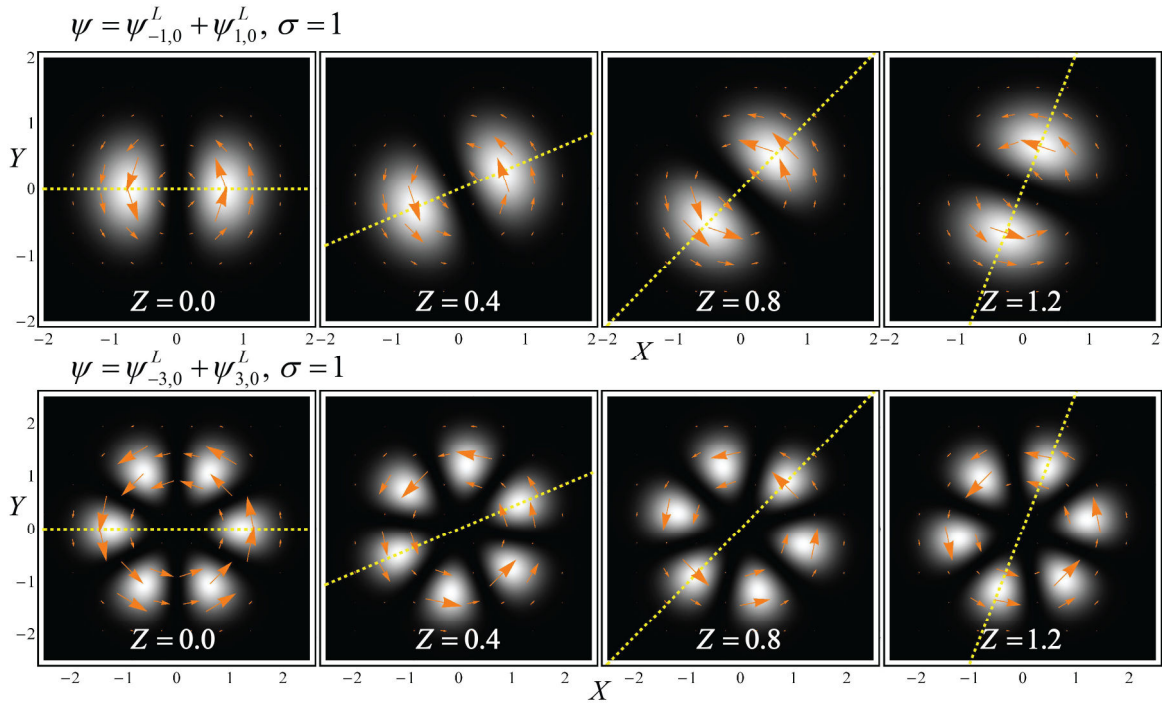

**Fig. 6.** Evolution of the "balanced" superposition (38) with $\ell = 1$ (upper panels) and $\ell = 3$ (lower panels) in the magnetic field with $\sigma = 1$. Probability densities and currents are shown. The rotation upon propagation along $z$, Eq. (39), corresponds to the Larmor frequency $|\Omega|$. The dimensionless coordinates $X = x/w_m$, $Y = y/w_m$, and $Z = z/z_m$ are used.

As another example, we consider an "OAM-unbalanced" superposition of two Landau modes (27) with the same number $n$ and vortex charges 0 and $\ell$:

$$\psi^{(2)} = \psi^L_{0,n} + a\psi^L_{\ell,n}, \tag{40}$$

where $a$ is some constant amplitude. Such superposition has a nonzero net canonical OAM, $\langle L_z \rangle \propto \ell$, and is characterized by a pattern with $|\ell|$ "off-axis vortices" [24,25,33,46,47] (see Fig. 7).

Landau modes with different $|\ell|$ involve the Gouy term in the difference of phases (36) and (37). Specifically, in the superposition (40), the mode $\psi^L_{\ell,n}$ acquires an additional phase $\Delta\Phi^{(2)}_{LZG} = -(\ell\sigma + |\ell|)\Phi_0$, as compared with the $\psi^L_{0,n}$ mode. From here it follows that the superposition (40) characterized by a parallel OAM and magnetic field, $\sigma\ell > 0$, will exhibit a rotation of its interference pattern by the angle

$$\Delta\varphi^>_{LZG} = 2\sigma z / z_m, \tag{41a}$$

whereas the superposition with an anti-parallel OAM and magnetic field, $\sigma\ell < 0$, will show no rotation at all:

$$\Delta\varphi^<_{LZG} = 0. \tag{41b}$$

Equation (41a) describes the rotation of the image with the *double-Larmor (i.e., cyclotron) frequency* $\omega_c = 2|\Omega|$, which corresponds to the cyclotron orbiting of the classical electron in a magnetic field. At the same time, no-rotation of Eq. (41b) should be associated with a classical trajectory *parallel* to the magnetic field (see below). As we pointed out above, in this case the kinetic quantum OAM becomes vortex-independent: $\langle\mathcal{L}^<_z\rangle = \hbar\sigma(2n+1)$. Examples of image rotations described by Eqs. (41) are shown in Figs. 7 and 8.

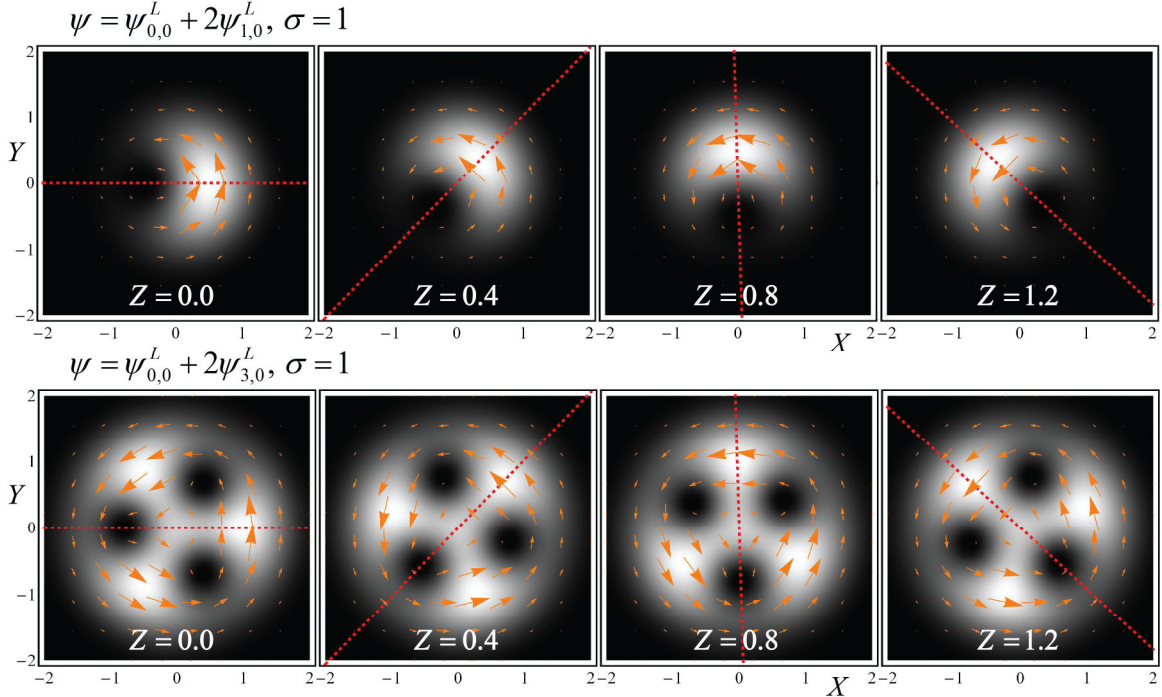

**Fig. 7.** Evolution of the superpositions (40) with $\ell = 1$ (upper panels) and $\ell = 3$ (lower panels) in the magnetic field with $\sigma = 1$. Probability densities and currents are shown. The rotation upon propagation along $z$, Eq. (41a), corresponds to the double-Larmor (i.e., cyclotron) frequency $\omega_c = 2|\Omega|$. The double rotation is caused by the addition of the transverse-current contributions from the vortex $\exp(i\ell\varphi)$ and from the vector-potential $A_\varphi$ (see Fig. 9 below). The dimensionless coordinates $X = x/w_m$, $Y = y/w_m$, and $Z = z/z_m$ are used.

The image rotation in an external magnetic field is well-known in electron microscopy [48]. There, it is associated with the classical cyclotron motion, and it is assumed that the rate of rotation always corresponds to the cyclotron frequency $\omega_c = 2|\Omega|$. However, from the above analysis it follows that the rotation of the quantum-electron interference patterns can be drastically different

and can demonstrate either Larmor-, cyclotron-, or even zero-frequency rotations, Eqs. (39) and (41). (For more complicated superpositions [24,49], the rotation rate can be between these frequencies and the evolution can be accompanied by deformations of the image.)

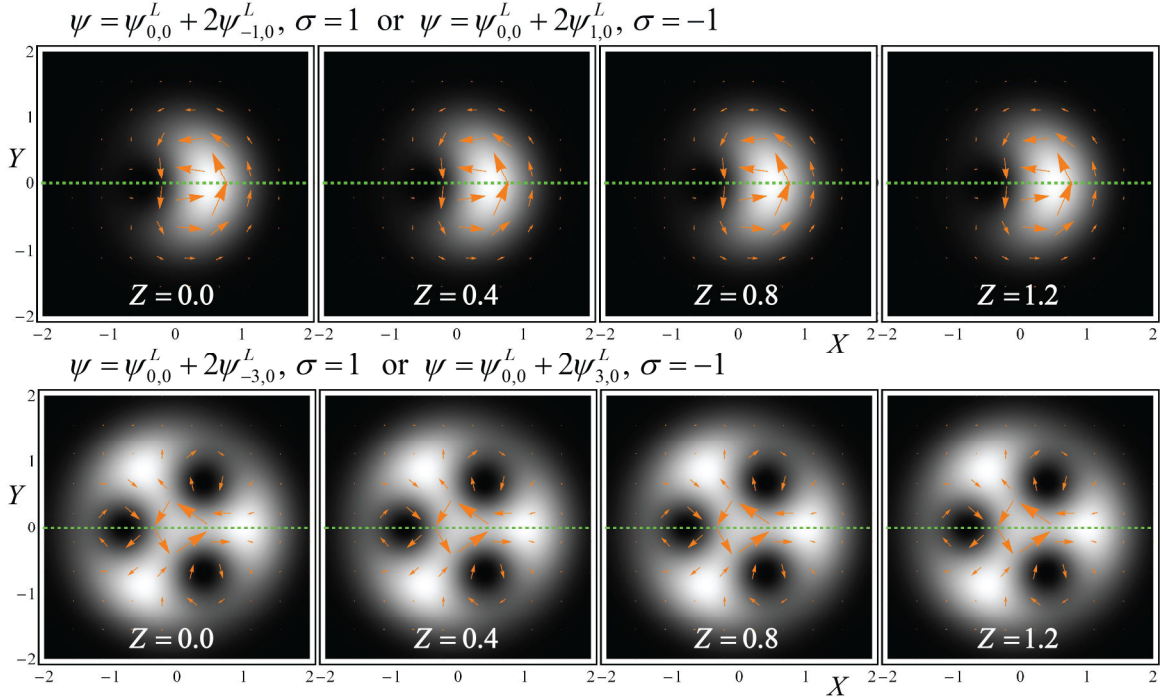

**Fig. 8.** Same as in Fig. 9, but for opposite vortex charges $\ell = -1$ and $\ell = -3$ (or opposite magnetic field, $\sigma = -1$). Propagation without rotation corresponds to Eq. (41b) and is explained by the cancelation of the transverse-current contributions from the vortex $\exp(i\ell\varphi)$ and from the vector-potential $A_\varphi$ (see Fig. 9 below).

Importantly, this rich rotational dynamics is fully consistent with the classical equations of motion in a magnetic field. According to the Ehrenfest theorem, the expectation values of the electron coordinates and momentum must obey the classical equations of motion:

$$\frac{d\langle \mathbf{r} \rangle}{dt} = \frac{\langle \boldsymbol{p} \rangle}{m}, \quad \frac{d\langle \boldsymbol{p} \rangle}{dt} = \frac{e}{m}\langle \boldsymbol{p} \rangle \times \mathbf{B}, \tag{42}$$

where $\langle \mathbf{r} \rangle = \langle \psi | \mathbf{r} | \psi \rangle / \langle \psi | \psi \rangle$ characterizes the coordinates of the *centroid* of the electron states, whereas $\langle \boldsymbol{p} \rangle = \langle \psi | \boldsymbol{p} | \psi \rangle / \langle \psi | \psi \rangle$ is the expectation value of the kinetic momentum $\hat{\boldsymbol{p}} = \hat{\mathbf{p}} - e\mathbf{A}$. For a single LG mode (27), we have

$$\langle \boldsymbol{p} \rangle = \langle \mathbf{p} \rangle = \hbar k_z \mathbf{e}_z, \quad \langle \mathbf{r} \rangle = z\, \mathbf{e}_z, \tag{43}$$

which corresponds to a rectilinear motion of the electron along the magnetic field with velocity $v = p_z / m$. A similar rectilinear motion of the centroid takes place for "balanced" superpositions (38). However, any superposition involving eigenmodes with OAM quantum number $\ell$ differing by $\pm 1$, will produce a transverse shift of the centroid and a simultaneous tilt of the beam in the orthogonal direction [33,47,50,51]. Such state acquires, e.g., $\langle x \rangle \neq 0$ and $\langle p_y \rangle \neq 0$ (which generates *extrinsic OAM* $\langle L_z^{\mathrm{ext}} \rangle = \langle x \rangle \langle p_y \rangle$ [41]), and the trajectory of the centroid becomes helical along the cyclotron orbit, Eqs. (42). Figure 9 shows examples of numerically-calculated centroid trajectories for the superpositions (40) with $\ell = \pm 1$ and $\sigma = 1$. Such superpositions represent off-axis vortices with shifted centroids. Strikingly, the centroids of similar superpositions with $\ell\sigma > 0$ and $\ell\sigma < 0$ demonstrate, respectively, the cyclotron orbiting and rectilinear propagation along the magnetic field. This is explained by the fact that in the case $\ell\sigma < 0$, the transverse component of the kinetic momentum vanishes, $\langle p_\perp \rangle = 0$, whereas $\langle p_\perp \rangle \neq 0$ in the $\ell\sigma > 0$ case. (At the same time, the canonical momentum has a non-zero transverse component, $\langle \mathbf{p}_\perp \rangle \neq 0$, in both cases.) Figure 9b shows that the addition and cancelation of the co- and counter-directing current contributions from the vortex $\exp(i\ell\varphi)$ and from the vector-potential $A_\varphi$ underpins such a remarkable behaviour.

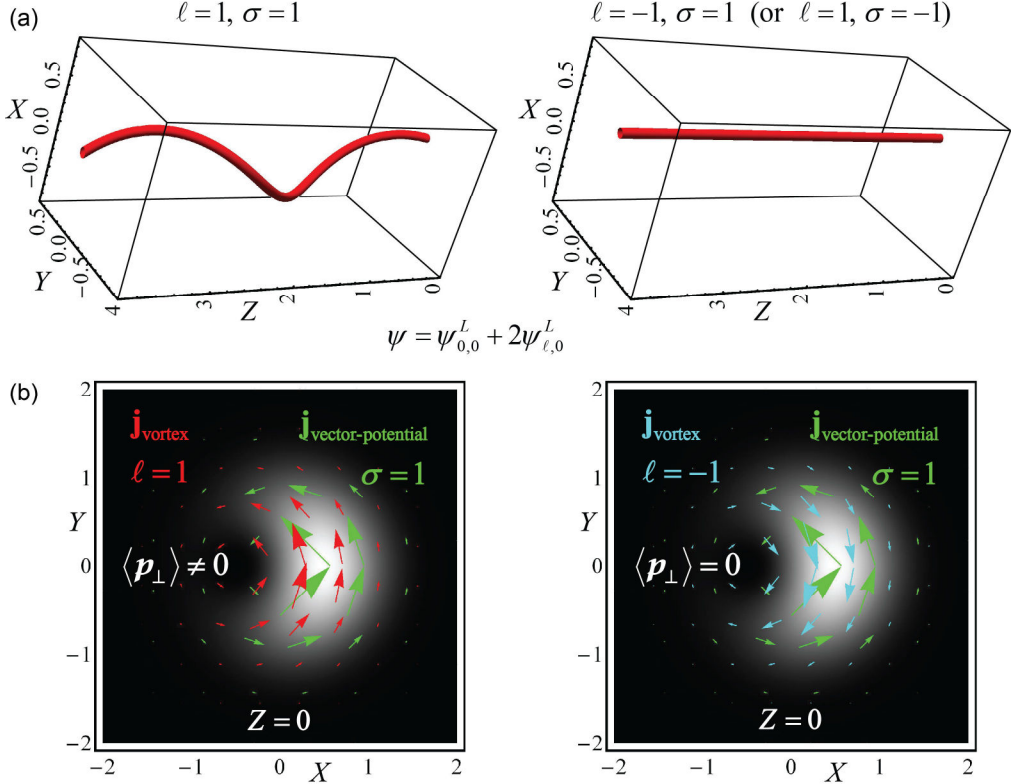

**Fig. 9.** (a) Trajectories of the centroids of superpositions (40) with $\ell = 1$ and $\ell = -1$ in a magnetic field with $\sigma = 1$. These trajectories satisfy the classical equations of motion (42) and correspond to evolutions shown in the upper panels of Figs. 7 and 8. (b) Transverse distributions of the probability densities $\rho$ and currents $\mathbf{j}_{\text{vortex}} = (\hbar/m)\text{Im}(\psi^*\nabla\psi)$ and $\mathbf{j}_{\text{vector-potential}} = -(e/m)\mathbf{A}\rho$, where $\mathbf{j}_{\text{vortex}} + \mathbf{j}_{\text{vector-potential}} = \mathbf{j}$, Eq. (15). In the cases $\ell\sigma > 0$ ($\ell\sigma < 0$), the addition (cancelation) of the vortex and vector-potential currents yields a non-zero (vanishing) transverse momentum $\langle p_\perp \rangle \neq 0$ ($\langle p_\perp \rangle = 0$). This explains the cyclotron orbiting and rectilinear propagation of the superpositions with $\ell\sigma > 0$ and $\ell\sigma < 0$, respectively.

In general, the rotation of the interference pattern and the orbiting of its centroid can be unrelated to each other. For instance, we calculated the evolution of the superposition $\psi = \psi_{-2,0}^L + \psi_{-1,0}^L + \psi_{1,0}^L + \psi_{2,0}^L$ (not shown here), and found that while the image follows the Larmor-frequency rotation due to the Zeeman effect for the opposite-$\ell$ modes, the centroid orbits with the cyclotron frequency. Naturally, such two-frequency evolution is accompanied by deformations of the interference pattern. In all cases, the interference of the LG modes (27) with the Landau-Zeeman-Gouy phases (36) and (37) underpins the evolution of any mixed electron state, whereas its centroid follows the classical equations of motion (42). In this manner, the quantum-interference picture complements the classical dynamics and significantly enriches the electron evolution in magnetic fields.

# 6. Conclusions

We have considered stationary electron vortex modes propagating in an external magnetic field. Although, formally, these modes (for the Landau and Aharonov-Bohm problems) have been known before, the discovery of *free* electron vortex beams gives a new twist to these classical problems. Indeed, traditionally, the Landau and Aharonov-Bohm modes were mostly considered as hidden (sometimes, auxiliary) states underlying the *transverse* transport effect in magnetic fields, and their OAM features were marginally investigated. In contrast, here we have considered *free propagation* of the vortex beam modes *along* the magnetic field (see Fig. 1) and paid particular attention to their canonical and kinetic OAM properties, as well as the probability density and current distributions. We have shown that it is natural to associate the Aharonov-Bohm and Landau states with the free Bessel and Laguerre-Gaussian vortex beams, which properties are modified by the presence of the magnetic field. This opens up a new avenue of investigations of these fundamental quantum states using electron vortex beams.

We have associated specific features of the Aharonov-Bohm and Landau modes with the coupling between the OAM from the vortex wave function and external magnetic field. In all cases, this breaks the symmetry between the vortex beams with opposite OAM values and brings about a rich variety of observable phenomena. The main manifestations of the OAM-magnetic field coupling in the Aharonov-Bohm and Landau problems are summarized in Table I.

| | Free-space analogues | Kinetic OAM | Probability density | Azimuthal current | Phase (dispersion) | Evolution (propagation) |
|---|---|---|---|---|---|---|
| AB states | Bessel beams | $\hbar(\ell-\alpha)$, *determines the beam radius, Eq. (24).* | Asymmetric w.r.t. $\ell \leftrightarrow -\ell$, $B \leftrightarrow -B$ | Asymmetric w.r.t. $\ell \leftrightarrow -\ell$, $B \leftrightarrow -B$ | $B$ and $\ell$ independent | Free propagation |
| Landau states | LG beams | $\hbar[\ell+\sigma(2n+|\ell|+1)]$ $=\hbar\sigma(2N+1)$, *determines Landau levels (29), (34) and phase (36), (37).* | Symmetric w.r.t. $\ell \leftrightarrow -\ell$, $B \leftrightarrow -B$ | Asymmetric w.r.t. $\ell \leftrightarrow -\ell$, $B \leftrightarrow -B$ | $B$ and $\ell$ dependent. Landau levels = Zeeman energy + Gouy phase | Image rotations: (i) Larmor, (ii) cyclotron, or (iii) zero rates |

**Table I.** Summary of the main features of the OAM-magnetic field coupling revealed in the Aharonov-Bohm (AB) and Landau stationary propagating modes. These modes represent the Bessel and LG beams in the presence of a magnetic flux line and uniform magnetic field, respectively.

For the Aharonov-Bohm Bessel states, the symmetry-breaking is sharply pronounced in the probability density distributions (see Fig. 4), while the phase and propagation characteristics are not affected by the magnetic flux line. This offers a novel approach to the Aharonov-Bohm effect, where the Dirac phase can be detected via measurements of the radius of the vortex beams propagating along the flux line [see Eq. (24)] [52].

In contrast to this, the probability density of the Landau LG modes is practically unaffected by the presence of the magnetic field, whereas the currents are strongly asymmetric and the resulting kinetic OAM is determined solely by the sign of the magnetic field and quantum numbers of the modes (see Fig. 5). Furthermore, the OAM-magnetic field interaction manifests itself in the dispersion relation for the propagation constant $k_z$, which underpins the Landau energy levels. We have revealed an intimate connection between the structure of the Landau levels, kinetic OAM, the Zeeman energy, and the Gouy phase, known in optics for diffracting LG beams. This brings about a rich propagation dynamics of superpositions of the LG modes in a magnetic field (Figs. 6-9). In particular, we have found that depending on the OAM properties of the superposition, it can rotate in a magnetic field with a rate corresponding to either the Larmor, cyclotron, or zero frequency.

The rotational interference effects described here deserve particular attention. One can find their analogues and manifestations in strikingly contrasting situations. Note that, according to the Larmor's theorem, a uniform magnetic field is equivalent to a rotation of the reference frame with the Larmor angular frequency. Then, subtracting the Larmor rotation from the dynamics shown in Figs. 6-9, one uncovers its similarity with free-space propagational dynamics. Specifically, the "balanced" flower-like superposition of Fig. 6 would undergo *no* rotation, whereas the "unbalanced" superpositions with "off-axis vortices" of Figs. 7 and 8 would demonstrate *opposite* rotations with the same angular velocities. This precisely coincides with the Gouy-phase behaviour observed in optical experiments with analogous superpositions of the diffracting LG beams [24–27]. Another remarkable analogy can be found in rotating trapped Bose-Einstein condensates (BEC) and superfluids (for a review, see [53]). There, the rotation of the trap can be associated with an external magnetic field, and the Landau problem naturally arises accompanied by the vortex excitations. According to our Fig. 5, the modes with minimal Landau energy $E_{\perp\min} = \hbar|\Omega|$ (and kinetic OAM $\langle \mathcal{L}_z \rangle_{\min} = \hbar$) are the LG modes with $n = 0$ and $\ell = 0, -1, -2, ...$ (assuming $\sigma = 1$). Superpositions (40) of these lowest-energy negative-vortex modes yields $|\ell|$ off-axis vortices with *no* rotation in the field, Fig. 8. This is the type of vortex excitations which is found to be stable in rotating Bose-Einstein condensates and superfluids, see the comparison in Fig. 10 [53–56].

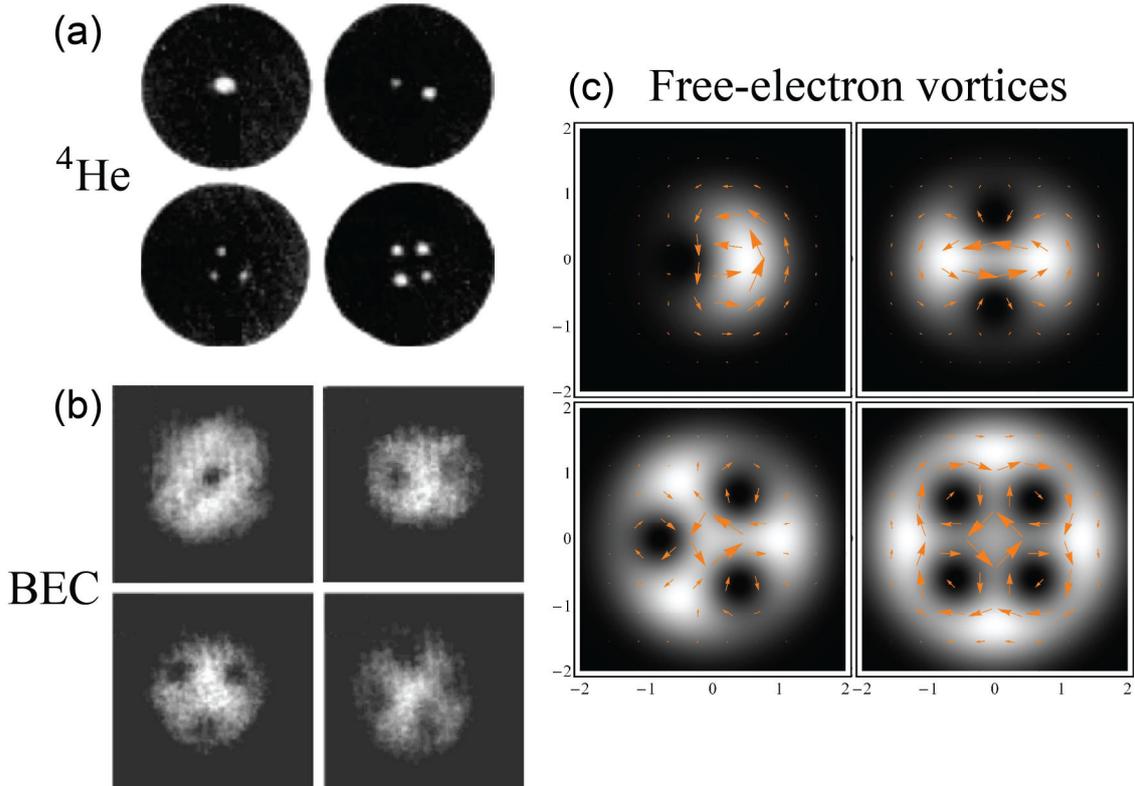

**Fig. 10.** Comparison of the stable vortex configurations in rotating $^4$He (a) (taken from [54]), and rotating trapped BEC (b) (taken from [55]), with nonrotating off-axis vortex superpositions of the lowest-energy Landau modes (c) (same as in Fig. 8 but with $\ell = -1, -2, -3, -4$).

We anticipate that the above theory will stimulate experimental observations of the Aharonov-Bohm and Landau modes using electron vortex beams. This enables to investigate novel manifestations of the Aharonov-Bohm effect as well as the Zeeman and Gouy phenomena, which underlie the Landau levels and corresponding kinetic OAM in a magnetic field. In this paper, we considered scalar electrons neglecting their vector (spin) properties. The effects of the spin evolution in the magnetic field can be readily added to the above picture because the spin-orbit coupling is negligible for paraxial electrons in a uniform field [10,12].

*Note added:* Another paper describing the OAM-dependent Zeeman phase and the Larmor-rotation effect for electron vortex beams appeared soon after submission of this work: C. Greenshields, R. L. Stamps, and S. Franke-Arnold, http://arxiv.org/abs/1204.4698v3. This paper terms the Zeeman-phase rotation as "Faraday rotation", in an analogy with the optical Faraday effect. It should be emphasized, however, that the optical Faraday effect is sensitive only to the polarization (spin) of light, but not to the OAM of the beam.

## Acknowledgements

We are grateful to Elena Ostrovskaya for drawing our attention to the analogy between the Landau problem and rotating trapped Bose-Einstein condensates. We also acknowledge support from the European Commission (Marie Curie Action), the Austrian Science Fund (project I543-N20), the European Research Council under the 7th Framework Program (FP7), ERC grant N°246791 – COUNTATOMS and ERC Starting Grant 278510 VORTEX, ARO, NSF grant No. 0726909, JSPS-RFBR contract No. 12-02-92100, Grant-in-Aid for Scientific Research (S), MEXT Kakenhi on Quantum Cybernetics, and the JSPS through its FIRST program.